\documentclass[showpacs,preprintnumbers,amsmath,amssymb,prl,twocolumn]{revtex4}
\usepackage{amsfonts}
\usepackage{mathrsfs}
\usepackage{graphicx}
\usepackage{dcolumn}
\usepackage{bm}

\begin{document}
\title{ Critical behaviors and local transformation properties of wave function}

\author{Jian Cui$^{\ast }$, Jun-Peng Cao$^{\ast }$, and Heng Fan }
\email{cuijian; junpencao; hfan@iphy.ac.cn} \affiliation{Beijing
National Laboratory for Condensed Matter Physics, Institute of
Physics, Chinese Academy of Sciences, Beijing 100190, People's
Republic of China}
\date{\today}

\begin{abstract}
We investigate crossing behavior of ground state entanglement
R\'{e}nyi entropies of quantum critical systems. We find a novel
property that the ground state in one quantum phase cannot be
locally transferred to that of another phase, that means a global
transformation is necessary. This also provides a clear evidence to
confirm the long standing expectation that entanglement R\'{e}nyi
entropy contains more information than entanglement von Neumann
entropy.  The method of studying crossing behavior of entanglement
R\'{e}nyi entropies can distinguish different quantum phases well.
We also study the excited states which still give interesting
results.
\end{abstract}

 \maketitle

\emph{Introduction}.--- Methods developed in quantum information
have proven to be very useful in studying the state of many-body
system \cite{REVIEWS OF MODERN PHYSICS}. At the same time experience
in condensed matter physics is helping in finding novel protocols
for quantum computation and communication. At the interface between
many-body system and quantum information, the analysis of
entanglement in quantum critical models has been attracting a great
deal of interests \cite{REVIEWS OF MODERN PHYSICS,Osterloh
nature,VidalKitaevprl,jiancui,Kit,Wen,xiaogangwen,Prlmc,Prlrandom}.

Entanglement, one intriguing feature of quantum theory and a main
resource for quantum information processing, is generally quantified
by entanglement von Neumann entropy (EvNE). It measures how closely
entangled the two subsystems are. It is shown that the behavior of
the critical EvNE is analogous to that of entropy in conformal field
theories which are for the quantum critical systems
\cite{VidalKitaevprl}. In addition, this EvNE contains the
topological entropy, a universal constant term and a topological
order, for a topological ordered state \cite{Kit,Wen}. A natural
generalization of EvNE is the entanglement R\'{e}nyi entropy (ERE)
which is believed to contain more information. However, it is shown
that ERE does not provide more information than EvNE for the case of
topological order \cite{xiaogangwen}. The question arising is: what
is the additional information in ERE not contained in EvNE and how
to use it in studying quantum critical phenomena? In this Letter, we
will provide an answer to this question.

For a pure bipartite state, $|\Psi _{AB}\rangle $,
the ERE is  defined with respect to a parameter $\alpha > 0$ as,
\begin{eqnarray}
S_{\alpha }(\rho _A)=\frac {1}{1-\alpha }\log _2[{\rm Tr}\rho _A^{\alpha }],
\end{eqnarray}
where $\rho _A$ is the reduced density operator of subsystem $A$ by
tracing out another subsystem $B$, $\rho _A={\rm Tr}_B(|\Psi _{AB}\rangle \langle \Psi _{AB}|)$.
Note that $\rho _B$ gives the same result.
In the limit, $\alpha \rightarrow 1$, ERE recovers the definition of EvNE,
$S_1(\rho _A)=S(\rho _A)\equiv -{\rm Tr}[\rho _A\log _2\rho _A]$.
Entanglement is invariant for local unitary transformations
in subsystems $A$ or $B$, so what matters in EvNE is the eigenvalue
spectrum of the reduced density operator. Actually the entanglement spectrum,  a redefinition
of eigenvalue spectrum, reveals much more information than EvNE, a single number \cite{LiHaldane}.
This motivated us still to consider an approach to extract the extra information from
ERE, a natural generalization of EvNE.

In studying quantum phase transitions by tools developed in quantum
information, generally the ground state properties, in particular
entanglement \cite{REVIEWS OF MODERN PHYSICS}
or fidelity
\cite{fidelity}, are studied. We may consider the EvNE, or similarly
concurrence, with various partitions.
The critical points can correspond to peaks or nonanalytic points
with different orders for quantities like EvNE, concurrence or
fidelity. Those methods, though approved to be powerful and
successful, have drawbacks. One drawback might be that there is no
unified standard to determine whether there exists a critical point
or not. Another drawback might be that the ground state property, in
particular from quantum information point of view, for different
quantum phases may not be completely revealed. Next in this Letter,
we will try to present a  powerful and novel approach different from
known methods, but without the drawbacks mentioned above.

\emph{Method}.---Entanglement does not increase under local
transformations. So one key property of a entanglement measure is
that it does not increase under local quantum operations and
classical communication (LOCC). Thus for pure bipartite quantum
states, it is only possible that a state with higher entanglement be
transferred by LOCC to a state with lower entanglement though it is
not always successful \cite{Nielsen}. The entanglement measure, such
as the well accepted EvNE, however, is not unique. In particular,
ERE with parameter $\alpha $ is also a entanglement measure.
Therefor, it is not surprising that the following case is possible:
For two bipartite pure states $|\psi _{AB}\rangle $ and $|\phi
_{AB}\rangle $ with reduced density operators denoted as
$\rho_{\psi_A}$ and $\rho_{\phi_A}$, when $\alpha =\alpha _1$, ERE
of $|\psi _{AB}\rangle $ is larger than that of $|\phi _{AB}\rangle
$, $S_{\alpha _1}(\rho_{\psi_A})>S_{\alpha _1}(\rho_{\phi_A})$;
while on the other hand when $\alpha =\alpha _2$, we have the
opposite direction, $S_{\alpha _2}(\rho_{\psi_A})<S_{\alpha
_2}(\rho_{\phi_A})$. That means neither state $|\psi _{AB}\rangle $
be transferred locally to state $|\phi _{AB}\rangle $ nor the
opposite transformation is possible, $|\psi _{AB}\rangle
\nleftrightarrow |\psi _{AB}\rangle $, since in each situation, one
entanglement measure would be increased by LOCC. In reversion, from
quantum information by ERE, it is shown that if the inequality holds
$S_{\alpha }(\rho_{\phi_A})\ge S_{\alpha }(\rho_{\phi_A})$ for all
$\alpha $, state $|\psi _{AB}\rangle $ can be transferred locally to
another state $|\phi _{AB}\rangle $ possibly assisted by a
``catalyst'' state \cite{jonathanplenio}. Thus the ERE provides a
necessary and sufficient condition for states local transformations
including intriguing ``catalyst'' case
\cite{Nielsen,jonathanplenio,necessarysufficient}.

In short, there are two different behaviors for EREs of two states
$|\psi _{AB}\rangle $ and $|\phi _{AB}\rangle $, see Fig.1: (left)
If there is no crossing, a state with higher entanglement can be
locally transferred to the lower entanglement one \cite{explain};
(right) If EREs are crossing, those two states can not be locally
transferred.

\begin{figure}
\includegraphics[height=3cm,width=\linewidth]{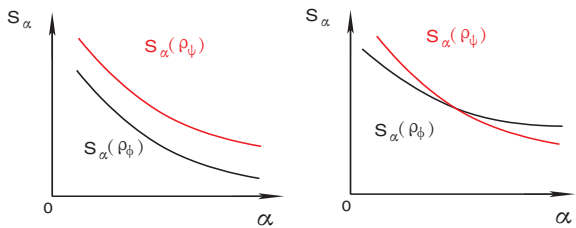}
\caption{\label{fig1}(Color online). Schematic of the method. ERE of
two bipartite states $|\psi \rangle $ and  $|\phi \rangle $ may have
two types of behavior: they are crossing or not. For non-crossing
case (left), $|\psi \rangle $ can be locally transferred to $|\phi
\rangle $. For crossing case (right), the two states cannot be
locally transferred to each other.}
\end{figure}

The above results can be applied to study the quantum critical
phenomena. We suppose when a quantum phase transition occurs, the
behavior of ground state ERE as well as the local transformation
property of the ground state wave function changes, and the
different quantum phases boundaries can be determined by the ERE. By
carefully analyzing the behavior of ERE, we find two cases. (i) In
some phases, the ground state EREs are crossing, while in other
phases, the ground state EREs are not crossing, please see the Table
I (left) and example model I. (ii) The ground states EREs do not
cross with others in the same phase, but they are crossing in the
different phases, please see the Table I (right) and example model
II. From the view of local transformation the above results are
explained as follows. (i) In some phases, ground states can not be
locally transferred into each other, i.e., a global transformation
is necessary; while in the other phases, ground states can be
locally transferred. (ii) The ground state can be transferred into
each other in the same phase with local transformation. However, the
ground state can not be transferred locally in the different phases.
These properties can be used to distinguish the quantum phases
transitions and the critical points can be found. We should notice
that the information about crossing can not be obtained from the
EvNE which is only the $\alpha =1$ case.

\begin{table}
\caption{\label{tab0} Crossing behaviors of the ground states EREs,
where crossed means EREs are crossed and N means the non-crossing.
The left table is for case (i) where the phase boundary can be
obtained along the diagonal elements. The right table corresponds to
the case (ii) where the phase boundary can be obtained with the help
of the anti-diagonal elements.}
\begin{normalsize}  
\begin{tabular}{|c|c|c|}
\hline EREs  &    phase I &    phase II \\
\hline phase I&   crossed & crossed\\
\hline phase II&  crossed &    N\\
\hline
\end{tabular}
\begin{tabular}{|c|c|c|}
\hline EREs  &    phase I &    phase II \\
\hline phase I&   N & crossed  \\
\hline phase II&  crossed & N\\
\hline
\end{tabular}
\end{normalsize}
\end{table}

\emph{Example model I: XY spin chain.}---The Hamiltonian of a 1D
spin-$1/2$ $XY$ chain takes the form of
\begin{eqnarray}
H=-\sum_i[(1+\gamma)\sigma_i^x\sigma_{i+1}^x+(1-\gamma)\sigma_i^y\sigma_{i+1}^y+h\sigma_i^z],
\end{eqnarray}
where $\sigma_i^{x,y,z}$ are Pauli matrices at site $i$, $\gamma$
and $h$ are coupling and field parameters. Here periodical boundary
condition is assumed. The phase diagram of $XY$ chain is presented
in Fig. 2. When the parameter $\gamma$ equals to one, the system (1)
degenerates into the 1D Ising model with transverse field,
$H_I=-\sum_i(\sigma_i^x\sigma_{i+1}^x+g\sigma_i^z),$ where $h=2g$.
It is well-known, a phase transition takes place at $g=1$ which is
gapless, while $g<1$ and $g>1$ are two gapped phases.

\begin{figure}
\includegraphics[height=4cm,width=5cm]{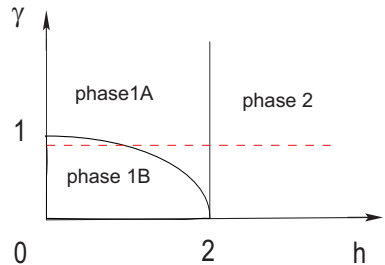}
\caption{\label{fig2}(Color online). Phase diagram of the $XY$
model. The phase transition from phase $1A$ or phase $1B$ to phase
$2$ is driven by the transverse magnetic field $h$, and the phase
transition from phase $1A$ to phase $1B$ is driven by the coupling
parameter $\gamma$. We focus on two cases: (a) $\gamma =1$, that is
the transverse Ising model where the critical point is $h=2$, and
(b) $\gamma=\sqrt 3/2$ (red dashed line) where the transition points
are $h=1$ and $h=2$. }
\end{figure}

In order to make a clear description of our method, we firstly study
the 1D Ising model with transverse field by the method of ERE. In
our method, we first find the ground state and partition it as two
parts $A:B$, $|G_{AB}\rangle $, we then calculate the ERE of this
ground state. In our numerical calculations, the total site $N$ is
taken to be $10$, and the chain is cut into two blocks with each $5$
sites respectively.

\begin{figure}
\includegraphics[height=3cm,width=\linewidth]{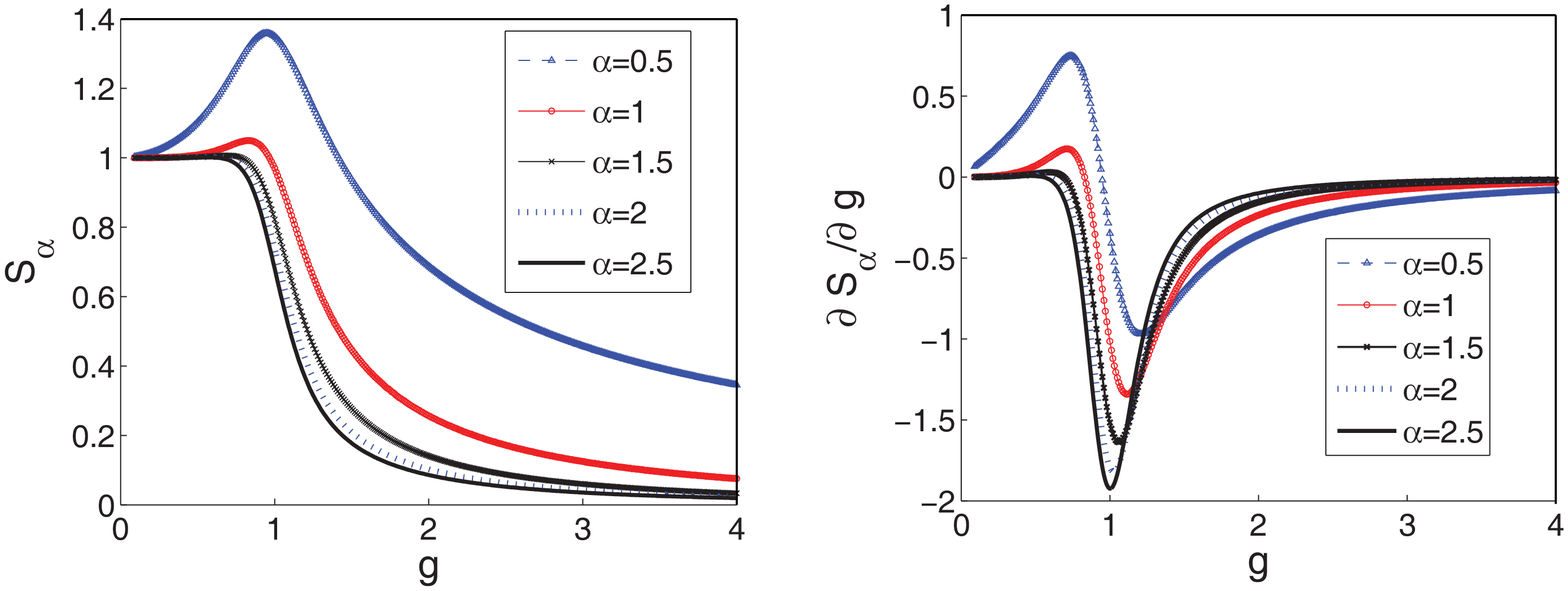}
\caption{\label{fig3}(Color online). EREs (left) and their first
derivative with respect to $g$ (right) for the ground state of Ising
model with transverse field. }
\end{figure}

\begin{figure}
\includegraphics[height=6cm,width=8cm]{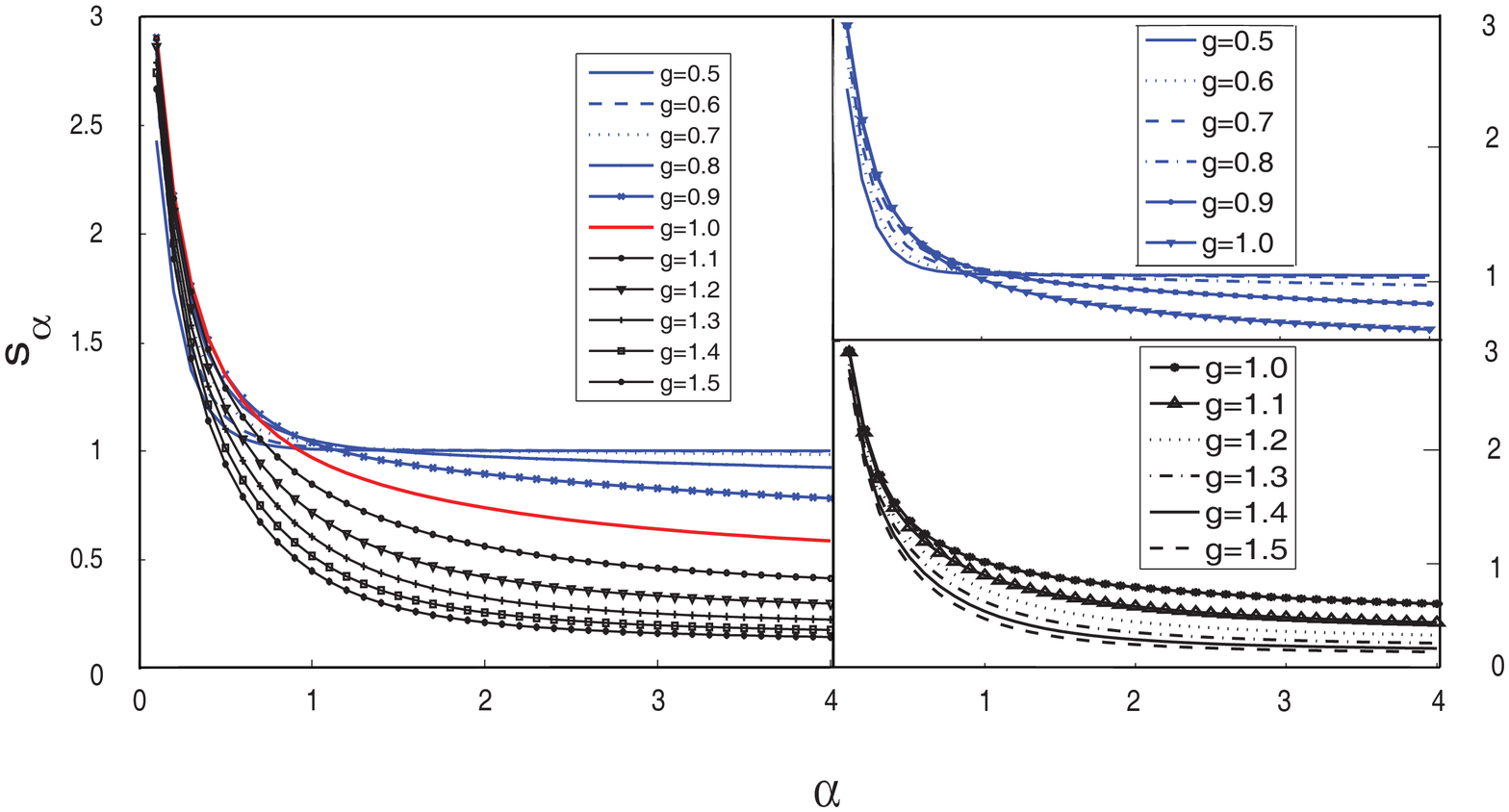}
\caption{\label{fig4}(Color online). The ground states EREs of
transverse Ising model with respect to $\alpha $. For $g < 1$, EREs
are crossing (right-up), for $g> 1$, EREs are non-crossing
(right-down) and $g=1$ is the critical point.}
\end{figure}

\begin{figure}
\includegraphics[height=3cm,width=\linewidth]{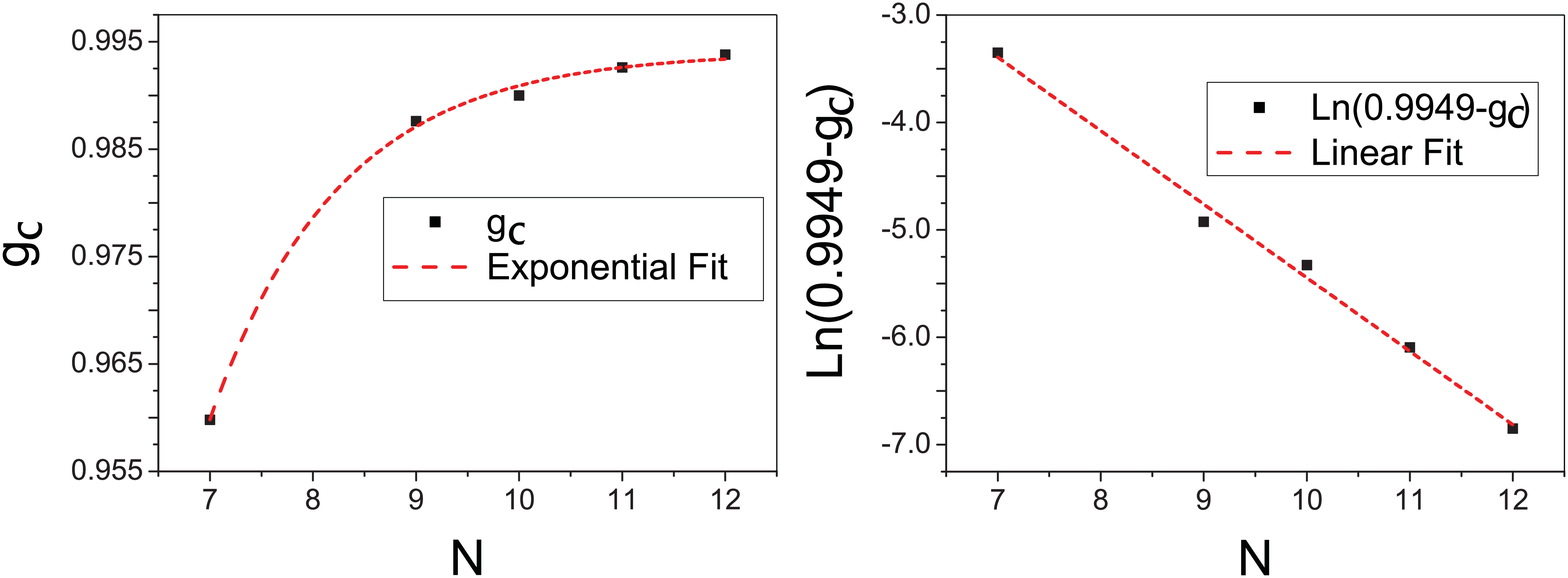}
\caption{\label{fig5}(Color online). The finite size scaling
behavior of the ground state EREs. }
\end{figure}

The results are presented in Figs. 3,4,5. We see that our method
works well for 1D transverse Ising model. The quantum phases are
clearly distinguished by different crossing behavior of EREs of the
ground state. Interestingly, for the two similar gapped phases,
their EREs are still different. That means their ground states local
transformation properties are different.

Let us go to a little bit detail.
We let parameter $g$ run from $0.5$ to $1.5$. There are three
distinct regions: region I (blue line) is the crossing region with
$g<1$, region III (black line) is the no crossing region with $g>1$,
and region II (red line) with $g$ around $1$ is formed by the
boundaries of I and III. Step by step, region II will be sharpened
to approach the critical point by narrowing the boundary interval.
First, the step size of $g$ is $0.1$ as shown in Fig. 4, and we can
fix that the phase transition happens at around $g=1$. Next, the
step size of $g$ is fixed to 0.01 to be more accurate and we let $g$
run from $0.94$ to $1.04$. For this case, the crossing figure is not
quite clear, so we use table to show where the crossing points are.
For example, in the first row of Table II, we find that $g=0.94$ get
crossed with $g=0.95, 0.96$,..., at $\alpha =0.6, 0.5$,.... For
region III, no crossing exists which is denoted as $N$. By Table II,
we find region II is $0.98\leq g\leq1.00$. We can go on
investigating this phase transition more accurately by the same
method and we list the result here: When step size of $g$ is
$0.001$, the critical region obtained by this method is $0.987\leq
g\leq0.989$; When step size of $g$ is $0.0001$, the critical region
is $0.9883\leq g\leq0.9885$. We can see that region II is sharpened
as parameter $g$ becomes more accurate. The finite size scaling
analysis is shown in Fig. 5. We see that the critical point obtained
by this method is $0.9949$, which is very close to the actual value.

\begin{table}
\caption{\label{tab1} The crossing points with parameter $g$ for the
transverse Ising model. For clearance, we cut the table into
separate parts, $g\le 0.98$, $g\ge 1$ and $0.98\le g\le 1$.}
\begin{scriptsize}  
\begin{tabular}{|c|c|c|c|c|c||c||c|c|c|c|c|}
\hline
  g & 0.94&0.95&0.96&0.97&0.98&0.99&1.00&1.01&1.02&1.03&1.04\\
  \hline

 0.94& N&  0.6 &0.5 &0.5 &0.5 &0.4 &0.4 &0.3 &0.3 &0.2 &N\\
  \hline
 0.95 &0.6 &N  &0.5 &0.5 &0.4 &0.4 &0.3 &0.3 &0.2& N& N\\
  \hline
  0.96 &0.5 &0.5 &N  &0.4 &0.4 &0.3 &0.3 &0.2  &N  &N
  &N\\
\hline
  0.97 &0.5 &0.5 &0.4 &N  &0.3 &0.3 &0.2   &N &N  &N&  N\\
  \hline
  0.98 &0.5 &0.4 &0.4 &0.3 &N  &0.2   &N  &N  &N  &N&  N\\
  \hline
  \hline
0.99 &0.4 &0.4 &0.3 &0.3 &0.2  &N &N &N &N& N &N\\
  \hline
  \hline
  1.00 &0.4 &0.3 &0.3 &0.2 &N    &N  &N  &N  &N  &N&
  N\\
  \hline
  1.01 &0.3 &0.3 &0.2&N  &N  &N    &N &N &N &N  &N\\
  \hline
  1.02 &0.3 &0.2 &N  &N   &N  &N  &N  &N  &N  &N
  &N\\
  \hline
  1.03 &0.2 &N  &N  &N  &N  &N  &N  &N  &N  &N
  &N\\
  \hline
  1.04 &N  &N  &N    &N  &N  &N  &N  &N  &N  &N
  &N\\
  \hline

\end{tabular}
\end{scriptsize}
\end{table}
Here we present also the interesting crossing phenomena of EREs
between ground state and the first excited state in different phases
for Ising model, see Fig. 6. In the ferromagnetic phase ($g<1$), the
ground state and the first excited state can not transfer locally,
while in the paramagnetic phase ($g>1$), the first excited state can
locally transfer to the ground state. This result generalizes our
previous ones and we can determine the zero temperature quantum
phase transition even using the finite temperature properties.

\begin{figure}
\includegraphics[height=3cm,width=\linewidth]{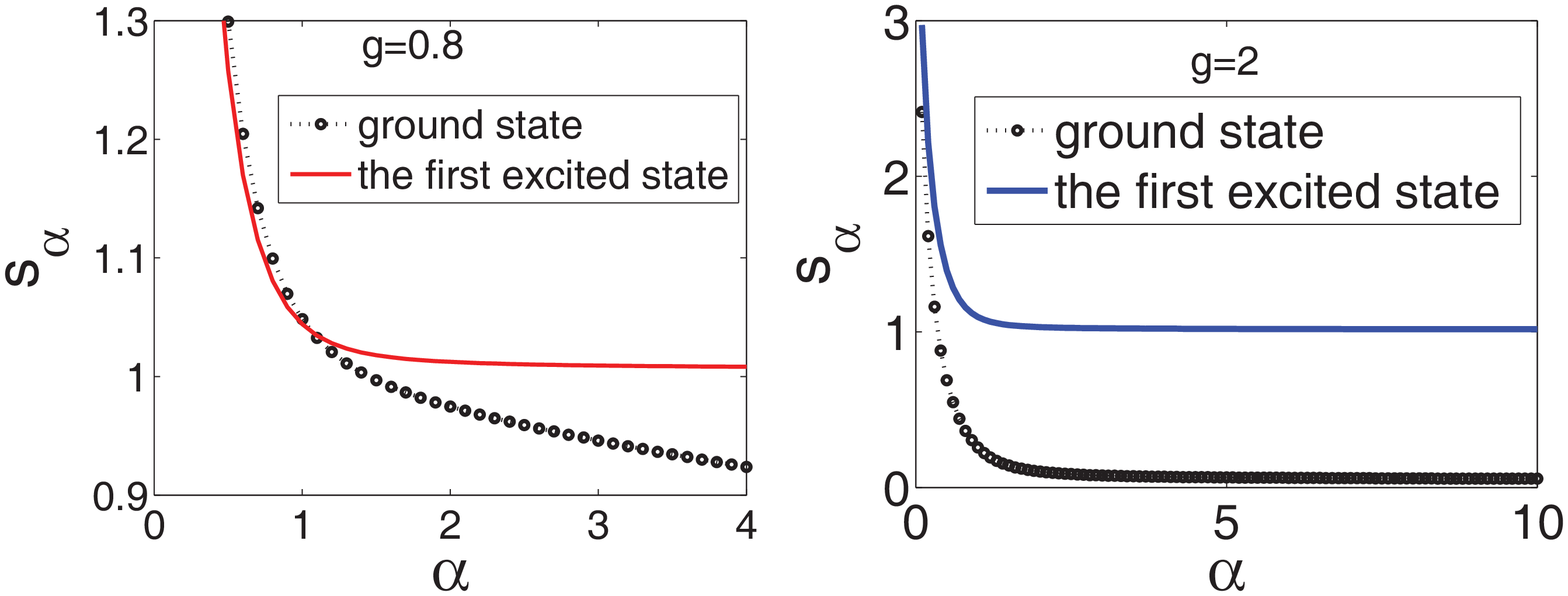}
\caption{\label{fig6}(Color online). The EREs of ground state and
first excited state of transverse Ising model. The EREs are crossing
in the phase $g<1$ and are not in the phase $g>1$.}
\end{figure}

Then, we consider the general case. Without loss of generality, we
consider the the red dashed line in Fig.2 where $\gamma=\sqrt3/2$.
Then the phase changes from $1B$ to $1A$ at $h=1$ and then from $1A$
to phase $2$ at $h=2$. We find that the ground state EREs of the
system with different magnetic field $h$ are crossed in phase $1B$,
not crossed in phase $1A$, and crossed again in phase $2$. Table III
gives a summary of the crossing results. The details of crossing
properties show that the phase transitions take place in regions
$0.999\leq h\leq1.000$ and $2.010\leq h\leq2.012$, which are very
close to the actual values $h=1$ and $h=2$.

\begin{table}
\caption{\label{tab4} The crossing points of ground state EREs of
the XY model near critical points.}
\begin{scriptsize}  
\begin{tabular}{|c|c|c|c|c|c|c|c|c|c|c|c|}
\hline h&   0.7 &0.8 &0.9 &1   &1.1 &1.2 &1.3     \\
\hline 0.7   &N&N&N&N&N&  0.5& 1.6  \\
\hline 0.8   &N&N&N&N&N&  1.6 &1.8 \\
\hline 0.9   &N&N&N&N&  1.4& 2   &1.9 \\
\hline 1     &N&N&N&N&  2.2 &2.2 &1.9 \\
\hline 1.1   &N&N  &1.4 &2.2 &N  &2.2 &1.9 \\
\hline 1.2   &  0.5 &1.6 &2   &2.2 &2.2 &N  &1.8\\
\hline 1.3  &1.6 &1.8 &1.9 &1.9 &1.9 &1.8 &N  \\
\hline
\end{tabular}
\begin{tabular}{|c|c|c|c|c|c|c|c|c|c|c|c|}
\hline h&   1.7 &1.8 &1.9 &2   &2.1 &2.2 &2.3   \\
\hline 1.7 &N  &0.9 &0.8 &0.7 &0.6 &0.4 &0.2 \\
\hline 1.8   &0.9 &N  &0.7 &0.6 &0.4 &0.3 &N \\
\hline 1.9  &0.8 &0.7 &N  &0.4 &0.2 &N&N\\
\hline 2    &0.7 &0.6 &0.4 &N&N&N&N\\
\hline 2.1  &0.6 &0.4 &0.2 &N&N&N&N\\
\hline 2.2 &0.4 &0.3 &N&N&N&N&N\\
\hline 2.3 &0.2 &N&N&N&N&N&N\\
\hline
\end{tabular}
\end{scriptsize}
\end{table}

\emph{Example model II: $XXZ$ spin chain.}---The Hamiltonian of
spin-$1/2$ $XXZ$ model is,
\begin{eqnarray}
H_{XXZ}=\sum_i\sigma_i^x\sigma_{i+1}^x+\sigma_i^y\sigma_{i+1}^y+\Delta
\sigma_i^z\sigma_{i+1}^z,
\end{eqnarray}
where $\Delta$ is the anisotropic parameter. There are two critical
points: $\Delta=-1$ corresponds to a first order phase transition,
$\Delta=1$ is a continuous phase transition. In particular phase
transition at $\Delta=1$ is a Kosterlitz-Thouless like transition,
the entanglements and their arbitrary order of derivatives are
analytic. We next try to identify the critical point $\Delta=1$ by
the ERE method.

Table IV shows the crossing points near $\Delta =1$. We can see that
each state in either region $\Delta \ge 1.0$ or $\Delta \le 1.0$
never cross with any of the states in the same region, but get
crossed with at least one state from the other region. The critical
region can be found to be $0.9\le \Delta \le 1.1$. By raising the
accuracy, this critical point can be found exactly. In all, our ERE
method also works well for the infinite order phase transition in
$XXZ$ spin chain.

\begin{table}
\caption{\label{tab2} The crossing points of the ground state EREs
of the $XXZ$ model. For clearance, we cut the table into separate
parts, $\Delta\le 0.9$, $\Delta\ge 1.1$ and $0.9\le \Delta\le 1.1$.}
\begin{scriptsize}  
\begin{tabular}{|c|c|c|c|c|c|c||c||c|c|c|c|c|c|}
\hline
  $\Delta$ &0.4& 0.5& 0.6 &0.7 &0.8 &0.9 &1.0   &1.1 &1.2 &1.3 &1.4
   &1.5&1.6\\
   \hline
    0.4 &N &N &N &N &N &N &N &N  &N &N &0.9 &0.5 &0.2\\
\hline
 0.5 &N&N &N &N &N &N &N &N &N  &N &0.6  &0.2&N\\
 \hline
 0.6 &N& N &N &N &N &N &N &N &N  &0.9 &0.2  &N&N\\
\hline
 0.7&N &N &N &N &N &N &N &N &N    &0.2 &N &N&N\\
 \hline
 0.8&N &N &N &N &N &N &N &N    &0.2 &N &N &N&N\\
\hline
 0.9 &N&N &N &N &N &N &N  &0.2 &N &N &N &N&N\\
\hline
 \hline
 1.0 &N&N &N &N &N &N &N &N &N  &N &N &N&N\\
\hline \hline
 1.1 &N&N &N &N &N  &0.2 &N &N &N &N &N &N&N\\
\hline
 1.2 &N&N &N &N  &0.2 &N &N &N &N &N &N &N&N\\
\hline
 1.3&N &N &0.9  &0.2 &N &N &N &N &N &N &N &N&N\\
\hline
 1.4 &0.9&0.6  &0.2   &N &N &N &N &N &N &N &N  &N&N\\
\hline
 1.5 &0.5&0.2   &N &N &N &N &N &N &N &N  &N &N&N\\
 \hline
  1.6 &0.2&N &N &N &N &N &N &N &N  &N &N &N&N\\
\hline
\end{tabular}
\end{scriptsize}
\end{table}

\emph{Summary.}---We propose a new method concerning about the local
transformation property of ground state to study quantum phase
transitions. Further, we have shown ERE contains more information
than EvNE in that by ERE we known deterministically whether two
ground states can be transferred locally to each other. As example
models, our method works well for 1D transverse Ising model and $XY$
spin chain. Interestingly, ground states similarly in two gapped
regions may possess different local transformation properties. Our
method also works well for the elusive critical point of $XXZ$ spin
chain. This simple and general method is worth (a) generalizing to
study finite temperature phase transitions (b) generalizing based on
the majorization scheme \cite{Nielsen} and (c) applying to other
systems.

We thank Zhi-Hao Xu and Zhao Liu for helpful discussions. This work
is supported by NSFC grants (10934010, 10974233, 10974247),
Knowledge Innovation Project of Chinese Academy of Sciences, and
``973'' program (2010CB922904, 2011CB921500, 2011CB921704).


\begin{thebibliography}{99}
\bibitem{explain}In this Letter, the local transformation includes the case of ``catalyst'', that means the
transformation may be assited by an ancillary entangled state.
\end{thebibliography}
\end{document}